\documentclass{article}
\usepackage{ijcai26}
\usepackage[T1]{fontenc}
\usepackage[utf8]{inputenc}

\usepackage{times}
\usepackage{microtype}
\usepackage{inconsolata}

\usepackage{soul}
\usepackage{url}
\usepackage[table]{xcolor}

\usepackage{graphicx}

\usepackage{amsmath,amssymb}
\usepackage{amsthm}
\usepackage{latexsym}

\usepackage{booktabs}
\usepackage{multirow}
\usepackage{adjustbox}

\usepackage{tikz}
\usepackage{pgf}

\usepackage{algorithm}
\usepackage{algpseudocode}

\usepackage{float}
\usepackage{etoolbox}
\usepackage{verbatim}

\usepackage[switch]{lineno}
\usepackage[hidelinks]{hyperref}

\title{Bridging the SEA Gap: An Initial Benchmark for Neural Audio Codec-Synthesized Speech Deepfakes in South-East Asian Languages}

\author{
Orchid Chetia Phukan\thanks{Corresponding Author}$^1$
\and
Girish .$^{1,2}$
\and
Mohd Mujtaba Akhtar$^{1,3}$
\and
Arun Balaji Buduru$^1$\\
\affiliations
$^1$IIIT-Delhi, India\\
$^2$UPES, India\\
$^3$VBSPU, India\\
\emails 
orchidp@iiitd.ac.in
}

\begin{document}

\maketitle

\begin{abstract}

\noindent Codecfakes (CFs) are a type of speech deepfakes generated through Audio Language Models (ALMs), with Neural Audio Codecs (NACs) forming the core mechanism for speech encoding and generation. CFs exhibit distributional characteristics that differ from vocoder-based deepfakes, causing detectors trained on vocoder data to generalize poorly to CFs detection. Although this has led to the development of CF detection benchmarks, existing resources are largely confined to English—and to a limited extent Chinese—leaving South-East Asian (SEA) languages unexplored. To bridge this gap, we introduce SEA-CF, the first large-scale benchmark for CF detection spanning multiple SEA languages, diverse speaker profiles, and a wide range of NAC architectures. SEA-CF is constructed by synthesizing publicly available real speech corpora. Our experiments show that state-of-the-art (SOTA) CF detectors trained on English-centric datasets fail to generalize to SEA speech due to language-specific phonetic structures, tonal variations, and rich prosodic diversity. We further conduct a comprehensive zero-shot and fine-tuned evaluation of recent SOTA ALMs on SEA-CF. Fine-tuning the ALMs improves performance, however, these are very large being impractical for real-world application due to their scale, particularly in low-resource and latency-constrained settings. To address this limitation, we propose a novel small-ALM, \textbf{\texttt{GARUDA}} tailored for CF detection, which delivers strong performance while remaining lightweight. Extensive evaluations demonstrate that the proposed Small-ALM outperforms strong end-to-end and ALM-based baselines, establishing a new, practical direction for robust CF detection in SEA languages and beyond.
\end{abstract}


\section{Introduction}

Recent advances in generative speech technologies, including text-to-speech (TTS) and voice conversion (VC), have enabled the synthesis of speech that closely matches natural human speech. In many cases, distinguishing machine-generated speech from real speech has become increasingly challenging, as modern systems can produce perceptually realistic voices across a wide range of languages. While these developments unlock significant opportunities in applications such as customer-facing services, healthcare, and entertainment, they also pose substantial security and societal risks. These include impersonation attacks, deepfake-enabled misinformation, and voice-based fraud in high-stakes domains such as digital banking and e-governance. To mitigate these threats, the speech research community has introduced a range of benchmarks for building models aimed at identifying synthetic speech~\cite{wu2015asvspoof,nautsch2021asvspoof,liu2023asvspoof,wang2024asvspoof}. Detection approaches have evolved steadily over time, showing consistent improvements over earlier methods. Early efforts relied on handcrafted acoustic features combined with classical machine learning techniques \cite{patel2015combining,patel2016cochlear}. This was later succeeded by deep learning–based models \cite{scardapane2017use,gomez2019light,chintha2020recurrent,jung2022aasist,muller2024mlaad}. Further, from beginning of the current decade, a substantial leap in speech deepfake detection performance was driven by the emergence of large pre-trained models (PTMs) \cite{wang2022investigating,tak2022automatic}. Consequently, the research community has extensively explored PTMs such as Whisper, XLS-R, Wav2vec2, WavLM etc. and proposing a wide range of novel detection frameworks that leverage these PTMs \cite{kawa23b_interspeech,hetero2024,huang-etal-2025-speechfake}.,
More recently, the community has began the usage of Audio Language Models (ALMs) such as Qwen2-Audio for speech deepfake detection. They have evaluated these ALMs either in a zero-shot or fine-tuning them for improved downstream performance \cite{gu2025allm4add,chuchra2026investigating}.\par

However, the majority of existing studies focus primarily on speech deepfakes generated using vocoder-based approaches, including traditional TTS and VC systems. This emphasis overlooks a rapidly emerging class of speech deepfakes i.e. Codecfakes (CFs) produced by ALMs, which differ fundamentally in their generation mechanisms and acoustic characteristics. ALMs rely on Neural Audio Codecs (NACs), including SoundStream, EnCodec, and DAC, as the backbone for speech encoding and generation. For instance, AudioLM employs SoundStream for discrete audio tokenization and synthesis \cite{borsos2023audiolm}. In response to this emerging threat, \cite{wu2024codecfake} and \cite{lu24f_interspeech} took the first steps by introducing dedicated CF benchmarks for the research community. Their studies demonstrated that state-of-the-art (SOTA) models trained on vocoder-based speech deepfake datasets fail to generalize when evaluated on CFs, highlighting the necessity of CF-specific detection approaches. These benchmarks have since fostered meaningful progress in CF detection \cite{zhu2025auddt}; however, they remain largely English-centric. While \cite{lu24f_interspeech} additionally includes Chinese as a secondary language, other languages remain entirely unexplored. This leaves a substantial gap in defense against CFs as previous research have shown that cross-lingual speech deepfake detection performance generally decreases \cite{ba2023transferring,hetero2024}. \par

In this work, we focus on CF detection in South-East Asian (SEA) languages. Our motivation stems from the fact that SEA, home to over 700 million people, is among the most linguistically and culturally diverse regions worldwide. Countries such as Indonesia, Thailand, Vietnam, the Philippines, Malaysia, Myanmar, Cambodia, Laos, and Singapore collectively encompasses languages spanning multiple families, including Austronesian (e.g., Malay, Indonesian, Tagalog), Austroasiatic (e.g., Vietnamese, Khmer), Tai–Kadai (e.g., Thai, Lao), and Sino-Tibetan (e.g., Burmese), each characterized by distinct phonetic, tonal, and prosodic structures. While \cite{wu2025seaspoof} introduced the first benchmark for speech deepfake detection in SEA languages, their study is limited to vocoder-based generation and does not consider CFs; moreover, the dataset has not been publicly released. To address these limitations, we introduce SEA-CF, the first publicly available and large-scale benchmark dedicated to CF detection in SEA languages, covering multiple languages, diverse speaker profiles, and a broad range of SOTA NAC architectures. SEA-CF is constructed by synthesizing publicly available real speech corpora using multiple NACs. Experimental results show that SOTA CF detectors trained on English-centric datasets fail to generalize to SEA speech, primarily due to linguistic differences. However, when jointly training previous benchmark CF dataset with SEA-CF, the performance improves, this shows the requirement of domain-specific training. We further perform comprehensive zero-shot and fine-tuned evaluations of recent SOTA ALMs. We evaluate these ALMs as previous research has shown their effectiveness compared to traditional end-to-end or PTM-based approaches for detecting speech deepfakes generated through vocoder-based approaches \cite{gu2025allm4add,chuchra2026investigating}. While fine-tuning improves detection performance, the large model sizes of these ALMs make them impractical for real-world applications, particularly in low-resource and latency-sensitive scenarios. Such real-world application is need of the hour for deepfake detection tools to be employed in real-world detection scenarios. To overcome this limitation, we propose a novel Small-ALM, \textbf{\texttt{GARUDA}}, tailored for CF detection. Extensive evaluations demonstrate that the proposed \textbf{\texttt{GARUDA}} consistently outperforms strong end-to-end and ALM-based baselines across both SEA-CF and previous CF detection benchmarks (achieving SOTA) while being lightweight and lesser inference time. \textbf{The main contributions are summarized as follows}: (i) We present SEA-CF, the first publicly available, large-scale benchmark for CF detection in SEA languages, encompassing multiple SEA languages, diverse speaker profiles, and a wide range of SOTA NACs. (ii) Extensive experiments on SEA-CF reveal that SOTA CF detectors trained on English-centric data generalize poorly to SEA languages. In contrast, joint training with SEA-CF and existing CF benchmarks improves performance, underscoring the necessity of in-domain training. (iii) We further perform comprehensive zero-shot and fine-tuned evaluations of recent SOTA ALMs. Our results reveal performance gains from fine-tuning. (iv) To address the real-world applicability challenges with large ALMs, we propose \textbf{\texttt{GARUDA}}, a novel Small ALM tailored for CF detection. Extensive evaluations show that \textbf{\texttt{GARUDA}} achieves SOTA performance across both SEA-CF and existing CF detection benchmarks while remaining lightweight and offering significantly reduced inference time. \noindent \textit{The SEA-CF dataset, code, and evaluation resources are available at: \footnote{\url{https://helixometry.github.io/SEACodecFake/}.}} \par

This work supports SDG 16 (Peace, Justice, and Strong Institutions) by strengthening defenses against codec-based speech deepfakes, especially for underrepresented SEA languages, and SDG 10 by reducing linguistic and regional gaps in deepfake detection. It also aligns with the “Leave No One Behind” principle by improving access to practical detection resources for low-resource language communities. The work was informed by academic collaboration and domain feedback from cybersecurity-oriented stakeholders, which helped shape the focus on lightweight and deployable CF detection systems. In particular, co-author Dr. Arun Balaji Buduru, a faculty member at IIIT-Delhi who is also associated with cybersecurity projects and fraud detection initiatives, provided valuable insights into the practical challenges faced in real-world deployments. These discussions highlighted several key requirements that directly influenced the design of our framework: (i) \textit{Language coverage gap}: Existing fraud detection systems are predominantly English-centric, leaving users of SEA languages vulnerable to emerging audio fraud threats. This motivated the development of the SEA-CF dataset to improve linguistic coverage and inclusivity. (ii) \textit{Low-latency deployment requirements}: Real-world fraud detection systems must operate under stringent latency and resource constraints. Consequently, we focused on a lightweight audio language model (less than 1B parameters) and optimized the framework to achieve an inference time of 1.21 seconds. (iii) \textit{Robustness to unseen NACs}: Fraudulent audio encountered in practice may originate from previously unseen synthesis pipelines. This motivated our explicit seen-versus-unseen codec evaluation protocol to assess cross-codec generalization and deployment robustness.

\section{Related Works}

\cite{wu2024codecfake} and \cite{lu24f_interspeech} made the research community to focus on the need for building models for CF detection as models trained on previous TTS and VC synthesis-based datasets fail here. \cite{wu2024codecfake} curated the CF dataset using the English VCTK corpus, while, \cite{lu24f_interspeech} developed CF corpora covering English and Chinese by leveraging VCTK and AISHELL-3 as source corpora. Both of these studies leverage AASIST-based modeling for CF detection. Subsequent work further expanded the CF detection landscape by incorporating a broader range of NAC families while continuing to rely on AASIST-based architectures as well as proposing an alternative CF detection strategy based on sharpness-aware minimization to improve robustness \cite{xie2025codecfake}. Despite these advances, existing CF datasets remain almost exclusively focused on high-resource languages, primarily English and Chinese, highlighting a clear lack of linguistic diversity. Notably, no prior work has systematically investigated CF detection in SEA languages. While \cite{cui2025whiadd} was the first to extend CF detection beyond English and Chinese by introducing a multilingual CF dataset, SEA languages were not included. In addition, \cite{wu2025seaspoof} proposed a dataset to facilitate speech deepfake detection in SEA languages; however, their work is limited to vocoder-based generation methods such as TTS and VC, and the dataset has not yet been made publicly available. Recent work has also begun extending synthetic speech forensics beyond closed-set binary detection, including Indic-language codec-fake detection~\cite{girish2026indic}, cross-lingual synthetic-speech attribution and emotional manipulation analysis~\cite{girish-etal-2025-towards}, open-set generator attribution for unseen synthesizers~\cite{akhtar-etal-2026-bridging}, geometry-aware generalization across diverse speech synthesis paradigms~\cite{sheth-etal-2025-curved}, and codec-fake detection in high-stakes healthcare settings~\cite{akhtar2026hcfd}. These studies further highlight the need for codec-specific, language-diverse, and efficient detection benchmarks. To bridge these gaps, we introduce the first large-scale and comprehensive benchmark dedicated to CF detection in SEA languages. Furthermore, we propose \textbf{\texttt{GARUDA}}, a novel Small-ALM designed for CF detection in SEA languages and beyond.

\section{SEA-CF}

\noindent In this section, we first describe the real-speech source corpora used to construct SEA-CF. We then present an overview of the SOTA NACs employed for CF generation, followed by a detailed description of the end-to-end data generation pipeline used to build the SEA-CF benchmark.

\noindent\textbf{Real-speech sources}: We adopt the same set of languages as SEA-Spoof~\cite{wu2025seaspoof}, namely Tamil, Hindi, Thai, Indonesian, Malay, and Vietnamese. For real-speech sources, we largely follow the datasets used in SEA-Spoof, including Mozilla Common Voice~\cite{ardila2020common} for multilingual coverage, GigaSpeech2~\cite{yang2025gigaspeech} for Indonesian, the Thai Dialect Corpus~\cite{suwanbandit2023thai} for Thai, and the VIVOS corpus~\cite{luong2016non} for Vietnamese. For Malay, we combine the Conversational Malay Speech Corpus\footnote{\url{https://magichub.com/datasets/malay-conversational-speech-corpus/}} with a curated Malaysian YouTube dataset processed using Whisper-Large\footnote{\url{https://huggingface.co/mesolitica/datasets}}. In contrast to SEA-Spoof, for Tamil and Hindi, we utilize the Indic-SUPERB corpus~\cite{javed2023indicsuperb}
, one of the largest publicly available datasets for these languages. Indic-SUPERB has also been employed in prior studies on speech deepfake detection for Tamil and Hindi, primarily focusing on vocoder-based TTS and VC generation approaches~\cite{sharma2025indicsynth}.

\noindent \textbf{Neural Audio Codecs (NACs)}: We adopt widely-used, publicly released NACs that are straightforward to reproduce and have been used by previous CF detection studies for generation for CFs \cite{lu24f_interspeech,wu2024codecfake}. \textbf{DAC} \cite{kumar2024high}: We employ DAC with 16 kHz, 24 kHz, and 44 kHz configurations. \textbf{Encodec} \cite{defossez2022high}: We use its 24 kHz and 48 kHz models. \noindent \textbf{SoundStream} \cite{zeghidour2021soundstream}: We use its 16 kHz version. \textbf{SpeechTokenizer} \cite{zhang2024speechtokenizer}: We utilize its default 16 kHz configuration. \textbf{FunCodec} \cite{du2024funcodec}: We adopt the official 16 kHz version. \textbf{AudioDec} \cite{wu2023audiodec}: We incorporate AudioDec with 28 kHz and 48 kHz variants. \textbf{SNAC} \cite{siuzdak2024snac}: We use SNAC operating at 24 kHz, 32 kHz, and 44 kHz. \textbf{MIMI} \cite{defossez2024moshi}: We utilize MIMI which operates at 24 kHz. We curate a repository compiling all the NACs used in our study and a easy to use pipeline for using them \footnote{\url{https://github.com/CodeVault-girish/Neural-Codecs}}.

\noindent \textbf{Generation Pipeline of SEA-CF}: To build SEA-CF, we employ a controlled resynthesis pipeline inspired by prior work on CF generation~\cite{wu2024codecfake}. CF samples are created by encoding and reconstructing real speech drawn from the SEA-CF source corpora using a diverse collection of NACs given above. Given an input speech waveform $s$, a NAC encoder $\mathcal{F}$ converts the signal into a discrete latent representation $q=\mathcal{F}(s)$. This representation is then passed through the corresponding decoder $\mathcal{G}$ to generate a reconstructed waveform $\hat{s}=\mathcal{G}(q)$, which is treated as the associated CF sample. This encoding--decoding process preserves the underlying linguistic content and speaker characteristics while introducing codec-specific artifacts unique to each NAC. Applying this procedure to every utterance across all selected NACs yields parallel datasets in which each real speech sample has a one-to-one CF counterpart for every codec configuration. To evaluate detection robustness, we define two experimental settings. (i) Seen setting: both training and evaluation involve CF samples generated using the same set of NACs (e.g., SNAC variants, DAC variants, Encodec, SoundStream, and SpeechTokenizer). (ii) Unseen setting: evaluation is conducted on CF samples synthesized using NACs not encountered during training (e.g., FunCodec variants, AudioDec variants, and MIMI), enabling assessment of cross-codec generalization. This pipeline is followed consistently across all SEA languages included in SEA-CF. We next describe the training, validation, and testing splits used in SEA-CF. For the (ii) Unseen evaluation setting, the test set exclusively consists of CF samples generated using NACs that are not observed during training. \textbf{Hindi and Tamil:} Since IndicSUPERB provides predefined training, validation, and test splits, we follow these splits directly for CF generation. IndicSUPERB further includes two test partitions, namely test-known and test-unknown. We use the test-known split for the (i) Seen evaluation setting and the test-unknown split for the (ii) Unseen evaluation setting. \textbf{All Other Languages:} For the remaining SEA languages, we follow the protocol in~\cite{wu2025seaspoof}, as the same real-speech sources are adopted. We randomly partition the data into training, validation, and testing sets using an 8:1:1 ratio. This partitioning is strictly preserved during CF generation to prevent data leakage.

\begin{figure}[!hbt]
\centering
    \includegraphics[width=0.79\linewidth]{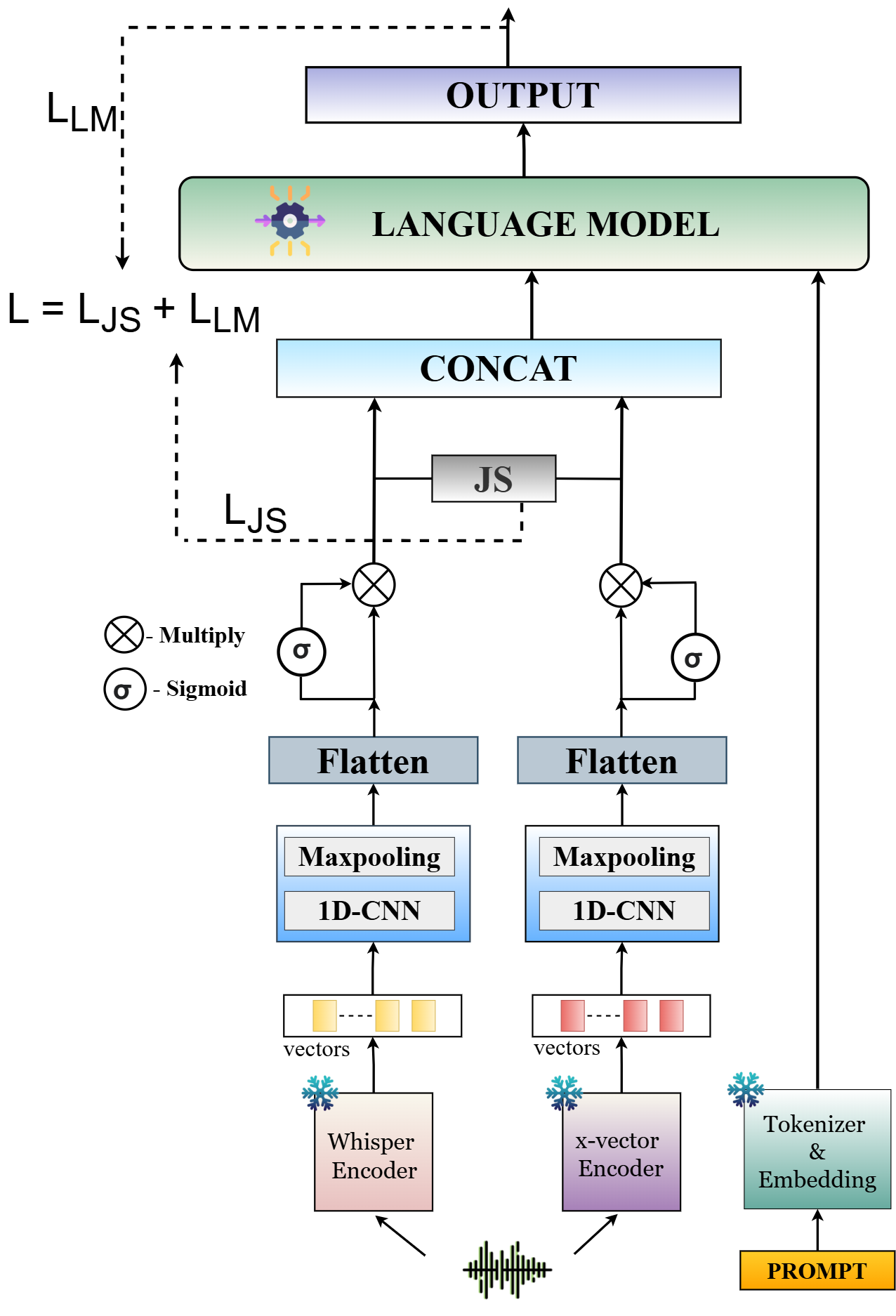}
    \caption{Proposed Framework: \textbf{\texttt{GARUDA}}}
    \label{archi}
\end{figure}


\section{Methodology}

\noindent In this section, we introduce our proposed framework, \textbf{\texttt{GARUDA}}. We first provide an overview of the overall architecture of \textbf{\texttt{GARUDA}}, followed by a detailed description of the audio encoders employed. Finally, we explain the language model (LM)-based decoder that enables high-level reasoning over fused audio representations. \textbf{\texttt{GARUDA}} adopts a dual-encoder design comprising Whisper and x-vector models to capture complementary aspects of speech. Whisper is leveraged to encode semantic and linguistic information, benefiting from its large-scale speech–text pretraining, while the x-vector encoder is employed to capture prosodic cues, such as pitch, tone attributing to its speaker recognition pretraining. Such dual-encoder architectures have been shown to be effective for speech deepfake detection in prior work, particularly through the fusion of semantic and prosodic cues \cite{hetero2024}. Similar dual-encoder fusion strategies have also been adopted in large ALMs for speech deepfake detection \cite{ranjan2025indicfake}. However, such approaches have not yet been explored in the context of CF detection within ALM-based frameworks. Furthermore, formulating speech deepfake detection as an audio question-answering task using ALMs has been shown to be effective in prior studies \cite{gu2025allm4add,li2025dfallm}, as it enables the model to perform context-aware reasoning and leverage the generative capabilities of ALMs beyond conventional discriminative classification.

\noindent\textbf{Audio Encoders}: \textbf{Whisper} \cite{radford2023robust}: It is a transformer-based encoder–decoder model trained in a multilingual, multi-task setting across 96 languages. In our framework, we utilize only the frozen encoder component to extract speech representations, producing a 512-dimensional embedding via average pooling. Owing to its training objective centered on automatic speech recognition, Whisper is particularly effective at encoding semantic and linguistic content in speech. We use the base version in our study which consists of 74M parameters. \textbf{x-vector} \cite{snyder2018x}: It is a time-delay neural network trained in a supervised manner for speaker recognition. The model is trained on a combination of the VoxCeleb1 and VoxCeleb2 datasets and contains approximately 4.2 million parameters. We use the frozen model and extract 512-dimension representations through average pooling.

\noindent \textbf{Language Model Decoder}: We adopt Qwen2-0.5B as the pre-trained decoder LM  backbone in our framework~\cite{team2024qwen2}. The use of a small LM (Less than 1B parameters) aligns with our motivation of designing a lightweight small-ALM that enables effective audio reasoning without the computational overhead of large-scale LLMs. This is due to the fact that the decoder LM generally remains heavyweight compared to its audio encoder counterparts in an ALM framework. Lighter the decoder LM lighter the ALM. As the smallest model in the Qwen2 family, Qwen2-0.5B offers an efficient yet capable architecture. Qwen2 LM family has demonstrated strong performance across speech and audio–language modeling tasks, including speech deepfake detection~\cite{li2025dfallm} and speech emotion recognition~\cite{su2025reasoning}. \newline
\noindent \textbf{Working Flow of \texttt{GARUDA}}: The archiecture of \textbf{\texttt{GARUDA}} is shown in Figure~\ref{archi}. Given a speech utterance, the objective of \textbf{\texttt{GARUDA}} is to determine whether the input audio is genuine or manipulated by generating a concise natural-language decision. To this end, we extract two complementary representations from the input: a semantic representation using Whisper and a prosodic representation using x-vector. These representations are subsequently processed through a convolutional module consisting of a 1D-CNN layer with a kernel size of three, followed by max pooling. After flattening, the features are passed through a sigmoid gating module that selectively filters the input representations, allowing only important information from the previous stage to passed through to the next layer. Suppose $x$ and $y$ represent the two feature vectors obtained from the two audio encoder branches. To encourage information-level consistency between these heterogeneous representations prior to fusion, we employ Jensen--Shannon (JS) divergence \cite{sutter2020multimodal} as a novel loss function. Since JS divergence is defined over probability distributions, we first transform the feature vectors into feature-wise distributions using temperature-scaled softmax normalization: $ p_x = \mathrm{softmax}\!\left(\frac{x}{\tau}\right) $ and $
p_y = \mathrm{softmax}\!\left(\frac{y}{\tau}\right) $ where $\tau$ denotes a temperature parameter controlling the smoothness of the distributions. We then define the mixture distribution as: $ m = \frac{1}{2}\left(p_x + p_y\right) $. The JS alignment loss is computed as: $\mathcal{L}_{\mathrm{JS}}
=
\frac{1}{2} \, \mathrm{KL}(p_x \| m)
+
\frac{1}{2} \, \mathrm{KL}(p_y \| m) $ where the Kullback--Leibler ($\mathrm{KL}$) divergence is defined as: $ \mathrm{KL}(p \| q) = \sum_{i} p_i \log \frac{p_i}{q_i} $.
For the JS loss, lower the value shows greater alignment and we will in later stage jointly optimize it with the language modeling loss. Following alignment, the two representations are concatenated and passed through a fully connected network of 90 neurons. The concatenated output is then projected into the LM embedding space and injected as a continuous prefix to the decoder LM. We use the prompt: \textit{Is the speech sample fake or real? Reply in one word "fake" or "real"}. Here the model is trained to predict only the word \textit{"fake"} or \textit{"real"} as the output. As our study doesn't focus on seeing the variance in performance due to the usage of different prompts, so we only experiment with a single prompt template. Finally, the decoder is trained using the standard language modeling objective $\mathcal{L}_{\mathrm{LM}}$. The language modeling objective minimizes the negative log-likelihood of the target sequence given a predicted sequence. The final training objective is given by: $ \mathcal{L}_{\text{total}} 
=
\mathcal{L}_{\mathrm{LM}}
+
\lambda \, \mathcal{L}_{\mathrm{JS}} $ where $\lambda$ controls the contribution of the JS alignment loss.

\begin{table}[!hbt]
\scriptsize
\centering
\setlength{\tabcolsep}{0.8pt}
\begin{tabular}{p{0.30\linewidth}|cc|cc|cc}
\toprule
\multirow{2}{*}{\textbf{Method}} &
\multicolumn{2}{c|}{\textbf{SEA-CF}} &
\multicolumn{2}{c|}{\textbf{CF \cite{lu2024codecfake}}} &
\multicolumn{2}{c}{\textbf{Avg}} \\
\cmidrule(lr){2-3}\cmidrule(lr){4-5}\cmidrule(lr){6-7}
& \textbf{ACC \(\uparrow\)} & \textbf{EER \(\downarrow\)} & \textbf{ACC \(\uparrow\)} & \textbf{EER \(\downarrow\)} & \textbf{ACC \(\uparrow\)} & \textbf{EER \(\downarrow\)} \\
\midrule
\multicolumn{1}{c}{} & \multicolumn{4}{c}{\textbf{Zero-shot evaluation of ALMs}} \\
\midrule
Qwen-Audio-Chat & 3.72 & 94.39 & 14.10 & 85.80 & 8.91 & 90.10 \\
Qwen-Audio-Base & 5.41 & 94.67 & 16.07 & 84.36 & 10.74 & 89.52 \\
Qwen2-Audio-Chat & 5.96 & 91.71 & 17.23 & 81.17 & 11.60 & 86.44 \\
Qwen2-Audio-Base & 8.41 & 91.53 & 19.48 & 80.47 & 13.95 & 86.00 \\
SeaLLMs-Audio-7B & 6.23 & 91.64 & 18.35 & 80.25 & 12.29 & 85.95 \\
\midrule

\multicolumn{1}{c}{} & \multicolumn{3}{c}{\textbf{End-to-end method}} \\
\midrule
AASIST & 86.98 & 15.74 & 93.09 & 8.16 & 90.04 & 11.95 \\
\midrule

\multicolumn{1}{c}{} & \multicolumn{3}{c}{\textbf{Pre-Trained Backbone}} \\
\midrule
Wh-LCCN & 87.69 & 15.22 & 94.41 & 7.63 & 91.05 & 11.43 \\
Wav2vec2-AASIST & 88.71 & 13.01 & 95.16 & 7.08 & 91.94 & 10.04 \\
MiO & 88.76 & 12.51 & 95.64 & 6.37 & 92.20 & 9.44 \\
\midrule

\multicolumn{1}{c}{} & \multicolumn{3}{c}{\textbf{FT}} \\
\midrule
SeaLLMs-Audio-7B & 88.74 & 9.64 & 90.75 & 6.96 & 89.75 & 8.30 \\
Qwen2-Audio-Base  & 93.88 & 6.95 & 95.06 & \cellcolor{yellow!20}4.21 & 94.47 & 5.58 \\
\midrule

\multicolumn{1}{c}{} & \multicolumn{3}{c}{\textbf{GARUDA}} \\
\midrule
Only Wh & 89.58 & 11.72 & 90.12 & 7.53 & 89.85 & 9.63 \\
Only XV & 90.99 & 10.43 & 93.15 & 7.28 & 92.07 & 8.86 \\
Wh+XV (Concat) & 91.56 & 9.17 & 94.88 & 7.16 & 93.22 & 8.17 \\
Wh+XV (CA) & 93.62 & 7.04 & 95.10 & 6.41 & 94.36 & 6.73 \\
Wh+XV (KL) & 90.83 & 9.31 & 93.46 & 6.68 & 92.15 & 8.00 \\
\textbf{\texttt{GARUDA}} &
94.37 & \cellcolor{yellow!20}6.26 &
97.00 & \cellcolor{yellow!20}4.19 &
95.69 & \cellcolor{yellow!20}5.23 \\
\midrule

\multicolumn{1}{c}{} & \multicolumn{3}{c}{\textbf{GARUDA-FT}} \\
\midrule
Only Wh & 91.87 & 10.31 & 92.80 & 6.43 & 92.34 & 8.37 \\
Only XV & 92.69 & 9.62 & 95.31 & 5.17 & 94.00 & 7.40 \\
Wh+XV (Concat) & 93.78 & 8.24 & 95.40 & 4.38 & 94.59 & 6.31 \\
Wh+XV (CA) &
\cellcolor{blue!20}96.07 & \cellcolor{blue!20}5.82 &
97.26 & 5.49 &
\cellcolor{yellow!20}96.67 & 5.66 \\
Wh+XV (KL) & 94.38 & 8.11 & 96.75 & 4.21  &95.57  & 6.16 \\
\textbf{\texttt{GARUDA-FT}} &
\cellcolor{green!40}\textbf{98.41} & \cellcolor{green!40}\textbf{2.78} &
\cellcolor{green!40}\textbf{99.36} & \cellcolor{green!40}\textbf{1.68} &
\cellcolor{green!40}\textbf{98.89} & \cellcolor{green!40}\textbf{2.23}
\\\midrule
\multicolumn{1}{c}{} & \multicolumn{3}{c}{\textbf{Training on Segments of Data}} \\
\midrule
Qwen2-Audio-Base-FT (75\%) & 89.51 & 8.67 & 93.24 & 5.92 & 91.38 & 7.30 \\
Qwen2-Audio-Base-FT (50\%) & 86.02 & 9.48 & 90.23 & 7.42 & 88.12 & 8.45 \\
Qwen2-Audio-Base-FT (25\%) & 82.93 & 10.73 & 88.07 & 8.76 & 85.50 & 9.75 \\
\textbf{\texttt{GARUDA}}(75\%) & 92.21 & 7.48 & 96.23 & 5.12 & 94.22 & 6.30 \\
\textbf{\texttt{GARUDA}}(50\%) & 91.48 & 8.27 & 95.17 & 6.56 & 93.33 & 7.42 \\
\textbf{\texttt{GARUDA}}(25\%) & 90.04 & 9.67 & 93.82 & 7.31 & 91.93 & 8.49 \\
\textbf{\texttt{GARUDA-FT}}(75\%) &
\cellcolor{yellow!20}95.12 & 6.34 &
\cellcolor{blue!20}98.43 & \cellcolor{blue!20}4.03 &
\cellcolor{blue!20}96.77 & \cellcolor{blue!20}5.19 \\
\textbf{\texttt{GARUDA-FT}}(50\%) & 93.56 & \cellcolor{yellow!20}6.89 & \cellcolor{yellow!20}98.24 & 5.47 & 95.90 & 6.18 \\
\textbf{\texttt{GARUDA-FT}}(25\%) & 92.04 & 8.29 & 96.26 & 6.19 & 94.15 & 7.24 \\
\bottomrule
\end{tabular}

\caption[Results on SeaCF and CodecFake]{
Results on SeaCF and CodecFake (CF)~\cite{lu2024codecfake}. 
All results are reported as percentages. 
Abbreviations: XV: x-vector; Wh: Whisper; FT: fine-tuning. 
\textbf{GARUDA-FT} denotes fine-tuning of the language-model decoder. 
KL denotes Kullback--Leibler divergence. 
These scores correspond to evaluation under the seen setting. 
The same abbreviations are used in Table~\ref{tab:seacf_codecfake_unseen}.
}
\label{table 1}
\end{table}

\section{Experiments}

\subsection{Training Details}
\textbf{\texttt{GARUDA}} is trained in a supervised setting on the combination of previous benchmark CF dataset \cite{lu2024codecfake} which contains English, Chinese and SEA-CF. We use \cite{lu2024codecfake} dataset as English and Chinese both are spoken in Singapore. We train on the combine training split as \cite{lu2024codecfake} have dedicated training, validation and testing split. Training of \textbf{\texttt{GARUDA}} is carried out in two formats: (i) Only training the projection module which consists of convolutional blocks, JS alignment loss, and the fully connected network (ii) training both the projection module as well as the LM decoder with LoRA. We use LoRA as fine-tuning full LM is cost intensive and LoRA provides a efficient way to fine-tune LMs. For (i), we train the models for 5 epochs with a batch size of 32 and learning rate of 1e-4. While for (ii), we train the models for 3 epochs with same batch size and learning rate of 1e-5. For LoRA, we set the rank to 8 and the scaling factor to 32. LoRA adapters are inserted into the query and value projection layers of the LM. For both the training formats, we apply dropout to prevent overfitting as well as use the same optimizer AdamW. We also kept the $\tau$ and $\lambda$ values to 0.5 and 0.4 respectively after initial experimentation on the validation set. For our experiments, we use A100 GPUs. 
\begin{table}[!hbt]
\scriptsize
\centering
\renewcommand{\arraystretch}{1.0}
\setlength{\tabcolsep}{1pt}
\begin{tabular}{p{0.30\linewidth}|cc|cc|cc}
\toprule
\multirow{2}{*}{\textbf{Method}} &
\multicolumn{2}{c|}{\textbf{SEA-CF}} &
\multicolumn{2}{c|}{\textbf{CF\cite{lu2024codecfake}}} &
\multicolumn{2}{c}{\textbf{Avg}} \\
\cmidrule(lr){2-3}\cmidrule(lr){4-5}\cmidrule(lr){6-7}
& \textbf{ACC \(\uparrow\)} & \textbf{EER \(\downarrow\)} & \textbf{ACC \(\uparrow\)} & \textbf{EER \(\downarrow\)} & \textbf{ACC \(\uparrow\)} & \textbf{EER \(\downarrow\)} \\
\midrule

MiO &
85.55 & 13.91 &
\cellcolor{yellow!20}93.44 & 7.77 &
88.50 & 10.84 \\
\midrule

Qwen2-Audio-Base-FT &
\cellcolor{yellow!20}92.08 & \cellcolor{yellow!20}8.15 &
93.26 & \cellcolor{yellow!20}5.42 &
\cellcolor{yellow!20}92.67 & \cellcolor{yellow!20}6.79 \\
\midrule

\textbf{\texttt{GARUDA}} &
\cellcolor{blue!20}92.97 & \cellcolor{blue!20}6.88 &
\cellcolor{blue!20}94.60 & \cellcolor{blue!20}5.71 &
\cellcolor{blue!20}93.79 & \cellcolor{blue!20}6.30 \\
\midrule

\textbf{\texttt{GARUDA-FT}} &
\cellcolor{green!40}\textbf{97.11} & \cellcolor{green!40}\textbf{3.17} &
\cellcolor{green!40}\textbf{98.06} & \cellcolor{green!40}\textbf{2.23} &
\cellcolor{green!40}\textbf{97.59} & \cellcolor{green!40}\textbf{2.70} \\
\bottomrule
\end{tabular}
\caption{Performance on Unseen Test Set}
\label{tab:seacf_codecfake_unseen}
\end{table}

\subsection{Experimental Results}

For testing, we test on Codecfake \cite{lu2024codecfake} and SEA-CF individually to quantify the performance. We use Accuracy (ACC) and Equal Error Rate (EER) as the evaluation metrics as used by previous speech deepfake detection research including CF detection studies \cite{wu2024codecfake,lu2024codecfake,hetero2024} (Table~\ref{table 1}). 

\noindent \textbf{SOTA model trained on CodecFake \cite{lu2024codecfake} and evaluated on SEA-CF}: We first access whether model trained on the prior benchmark \cite{lu2024codecfake} which consists of English and Chinese--can generalize to SEA-CF without any SEA-specific training. We use AASIST as the modeling architecture as it was shown to be effective by both \cite{lu2024codecfake} and \cite{wu2024codecfake} and represents SOTA model. Training and evaluation on CodecFake \cite{lu2024codecfake} dataset gives us around $94.08\%$ ACC and $6.76\%$ EER. However, when evaluated on SEA-CF, we get $70.65\%$ ACC $28.13\%$ EER, thus showing extreme degradation in performance and the need for in-domain training on SEA-CF. 

\noindent \textbf{Zero-shot evaluation of SOTA ALMs}: We evaluate SOTA ALMs in a strictly zero-shot setting by prompting them to classify each utterance with the same prompt as used for \textbf{\texttt{GARUDA}}. As shown in Table~\ref{table 1}, zero-shot ALMs are generally unreliable for CF detection across both SEA-CF and CodecFake \cite{lu2024codecfake}. We consider Qwen-audio and Qwen2-audio families of large ALMs as considered by \cite{gu2025allm4add} which is first study of evaluating ALMs for speech deepfake detection excluding CF detection. We also include SeaLLMs-Audio-7B \cite{liu2025seallms} as this ALM is fine-tuned on SEA languages. The results are presented in Table \ref{table 1}. Qwen2-Audio-Base performs the best across all ALMs. However, the performance across all the ALMs remains quite low. SeaLLMs-Audio-7B despite being trained especially for SEA languages performs lower than Qwen2-Audio-Base. This performance of Qwen2-Audio-Base is also supported by \cite{gu2025allm4add}. So, we keep Qwen2-Audio-Base as the SOTA ALM architecture for comparison in our succedding experiments.

\noindent \textbf{In-domain Training: Classifier Baselines and SOTA ALMs Finetuning}: We present the results with training done on combination of Codecfake \cite{lu2024codecfake} and SEA-CF and evaluated on SEA-CF and Codecfake \cite{lu2024codecfake}. The results are given in Table \ref{table 1}. We consider several baselines including End-to-end (AASIST \cite{jung2022aasist})and Pre-Trained Backbone (Whisper-LCNN~\cite{kawa23b_interspeech}, Wav2vec2-AASIST~\cite{lu2024codecfake}, MiO~\cite{hetero2024}) approaches. These are traditional classifier-based baselines. MiO performs the best showing the effectiveness of dual-encoder fusion due to the emergence of complementary behavior amongst them. Then we present the results of finetuning large ALMs, we fine-tune Qwen2-Audio-Base and SeaLLMs-Audio-7B with LoRA with the same configuration as used for \textbf{\texttt{GARUDA}} for fair comparison. We consider Qwen2-Audio-Base and SeaLLMs-Audio-7B for finetuning experiments as they have shown relatively better performance than other large ALMs in zero-shot setting. We see that in-domain training on mixture of SEA-CF and CodecFake \cite{lu2024codecfake} generally improves performance over their zero-shot performances as well as traditional classifier-based baselines. These shows the importance of in-domain training as well as effectiveness of forming the deepfake detection task as a audio question answering task.

\noindent \textbf{In-domain Training: \texttt{GARUDA}}: We first report results under training setting (i), where only the projection modules are trained. As shown in Table~\ref{table 1} under the \textbf{\texttt{GARUDA}} section, \textbf{\texttt{GARUDA}} consistently outperforms fine-tuned Qwen2-Audio-Base and SeaLLMs-Audio-7B, despite being significantly more lightweight and without any decoder fine-tuning. Notably, both Qwen2-Audio-Base and SeaLLMs-Audio-7B are built upon 7B-parameter LLM backbones. These results highlight the effectiveness of dual-encoder fusion driven by complementary audio representations and the proposed JS alignment loss. Furthermore, our findings reinforce prior observations that audio encoders constitute the primary performance bottleneck in ALMs, while the choice of the LM decoder plays a comparatively less critical role than the effective selection and fusion of audio encoders~\cite{li2025dfallm}. We additionally conduct ablation studies by selectively removing components of \textbf{\texttt{GARUDA}} to assess their individual contributions. Only Wh and Only XV correspond to configurations where a single audio encoder (Whisper or x-vector) is used within the \textbf{\texttt{GARUDA}} framework while keeping the remaining modeling unchanged. Wh+XV (Concat), Wh+XV (CA), Wh+XV (KL) denote fusion via simple concatenation without JS alignment loss, cross-attention and KL divergence alignment, respectively. While these variants outperform traditional classifier-based baselines, their performance remains inferior to the fine-tuned Qwen2-Audio-Base. This observation underscores the importance of effective audio encoder fusion for achieving optimal performance as shown by \textbf{\texttt{GARUDA}}. Secondly, we report results under training setting (ii), where the LM decoder is fine-tuned. The corresponding results are summarized in Table~\ref{table 1} under the \textbf{\texttt{GARUDA-FT}} section. Compared to training setting (i), we observe a consistent performance improvement, as fine-tuning the decoder mitigates hallucination effects and enables better alignment of the model with the downstream task.

\noindent \textbf{Training on Segments of Data}: We further conduct experiments by training the models on subsets of the training data, following the protocol adopted in~\cite{gu2025allm4add}. The results are reported in Table~\ref{table 1} under the Training on Segments of Data section. We evaluate Qwen2-Audio-Base-FT (fine-tuned Qwen2-Audio-Base), \textbf{\texttt{GARUDA}} under training setting (i), and \textbf{\texttt{GARUDA-FT}} under training setting (ii). Specifically, models are trained using 75\%, 50\%, and 25\% of the training data and evaluated on the full test sets of SEA-CF and CodecFake~\cite{lu2024codecfake}. Across all data regimes, \textbf{\texttt{GARUDA-FT}} consistently achieves the best performance, demonstrating strong robustness under limited-data training conditions.

\noindent \textbf{Evaluation on Unseen Test Set}: All results discussed above correspond to evaluation setting (i), where the test data is drawn from seen conditions. We further evaluate our models under evaluation setting (ii), which considers unseen test conditions. The corresponding results are reported in Table~\ref{tab:seacf_codecfake_unseen}. Using models trained on the full training set, \textbf{\texttt{GARUDA}} remains consistently strong under held-out conditions, outperforming both traditional classifier-based baselines and the ALM baseline.

\noindent \textbf{Summary}: Overall, our experimental results demonstrate that fusing complementary audio encoders and fine-tuning the LM decoder leads to superior performance compared to fine-tuning the decoder with suboptimal audio encoders (e.g., Qwen2-Audio-Base and SeaLLMs-Audio-7B) or training without decoder fine-tuning.

    \noindent \textbf{Lightweight and Efficiency Analysis}: We perform an efficiency analysis of \textbf{\texttt{GARUDA}}. \textbf{\texttt{GARUDA}} consists of Qwen2-0.5B LM decoder. Their audio encoders, Projection and LoRA parameters sums up to be approx 110M so, the total parameters still less than 1B far less then Qwen2-Audio-Base and SeaLLMs-Audio-7B. Also, to quantify, JS adds no parameters on top of simple concatenation based fusion as it is a loss function and no trainable parameters. Further, we give an latency analysis over the whole seen training (test) set, Qwen2-Audio-Base (Finetuned) incurs an average inference time of 12.32 seconds whereas \textbf{\texttt{GARUDA-FT}} only 1.21s. To sum up, \textbf{\texttt{GARUDA}} is effective framework for CF detection in SEA-CF and beyond while being extremely efficient in both model size as well as inference time. 

\noindent \textbf{Comparison to previous SOTA models on the CodecFake benchmark~\cite{lu2024codecfake}}: Since AASIST, Wav2vec2-AASIST, and MiO represent prior SOTA approaches for speech deepfake detection, including CF detection, our small-ALM framework, \textbf{\texttt{GARUDA}}, establishes a new SOTA on the CodecFake dataset~\cite{lu2024codecfake}.

\noindent \textbf{Statistical significance Test}: We evaluate the statistical reliability of the observed performance gains using a paired McNemar’s test on the same test sets employed in prior audio deepfake detection studies~\cite{batra2025melody}. Specifically, we compare \textbf{\texttt{GARUDA-FT}} against the strongest baselines, MiO and Qwen2-Audio-Base (Finetuned), using paired predictions on identical evaluation samples. \textbf{\texttt{GARUDA-FT}} shows statistically significant improvements over both MiO and Qwen2-Audio-Base (Finetuned) across SEA-CF and CodecFake~\cite{lu2024codecfake} benchmarks. The reported p-values correspond to pairwise McNemar tests comparing \textbf{\texttt{GARUDA-FT}} against each baseline: 0.0047 (vs MiO) and 0.00021 (vs Qwen2-Audio-Base (Finetuned)) on SEA-CF, and 0.0003 (vs MiO) and 0.00019 (vs Qwen2-Audio-Base (Finetuned)) on CodecFake~\cite{lu2024codecfake}, all well below the 0.05 threshold. This indicates that the observed performance gains of \textbf{\texttt{GARUDA-FT}} are consistent and not due to random variation.


\section{Conclusion}

In this work, we introduce SEA-CF, the first large-scale benchmark specifically designed for CF detection in SEA languages. SEA-CF covers multiple languages, diverse speaker profiles, and a broad spectrum of SOTA NAC architectures. Our experiments reveal that existing SOTA CF detectors trained predominantly on English-centric datasets exhibit poor generalization to SEA speech, underscoring the necessity of in-domain training for SEA languages. To address this gap, we propose \textbf{\texttt{GARUDA}}, a novel small-ALM that achieves SOTA performance compared to both traditional classifier-based CF detectors and large ALM baselines such as Qwen2-Audio-Base and SeaLLMs-Audio-7B. Despite being significantly more lightweight, \textbf{\texttt{GARUDA}} consistently delivers superior performance while requiring substantially fewer parameters and lower inference latency. Overall, our work establishes a strong foundation for future research on CF detection in low-resource SEA languages and beyond. Furthermore, our work also presents an efficient and practical framework for real-world applicability.

\noindent \textbf{Limitations and Future Work}: Despite being the first effort toward constructing a CF benchmark for SEA languages, our work has several limitations. First, SEA-CF does not yet cover all SEA languages; we plan to extend the benchmark to additional languages in future work. Second, as the SEA-Spoof dataset has not yet been released, our evaluation is currently restricted to the available benchmarks. Once SEA-Spoof becomes publicly available, we aim to develop more generalized models capable of detecting speech deepfakes across diverse generators. Future work will also explore reasoning-centric ALMs for CF detection, with a focus on equipping detectors with natural language explanation capabilities that can articulate why a given speech sample is classified as real or fake, thereby facilitating easier interpretation and adoption by end users.

\section*{Disclosure for Use of AI Assistants}
AI assistants were used solely for language refinement. All technical contributions, experimental design, modeling, dataset construction, and analysis were carried out by the authors.

\section*{Authors Contribution}
Orchid Chetia Phukan, Girish, and Mohd Mujtaba Akhtar contributed equally as first co-authors. 
\bibliographystyle{named}
\bibliography{ijcai26}

\end{document}